# The Rise of Data-Driven Microscopy powered by Machine Learning

Leonor Morgado[1], Estibaliz Gómez-de-Mariscal[1], Hannah S. Heil[1]✉, and Ricardo Henriques[1,2]✉

[1]Instituto Gulbenkian de Ciência, Oeiras, Portugal
[2]MRC-Laboratory for Molecular Cell Biology. University College London, London, United Kingdom

Optical microscopy is an indispensable tool in life sciences research, but conventional techniques require compromises between imaging parameters like speed, resolution, field-of-view, and phototoxicity. To overcome these limitations, data-driven microscopes incorporate feedback loops between data acquisition and analysis. This review overviews how machine learning enables automated image analysis to optimise microscopy in real-time. We first introduce key data-driven microscopy concepts and machine learning methods relevant to microscopy image analysis. Subsequently, we highlight pioneering works and recent advances in integrating machine learning into microscopy acquisition workflows, including optimising illumination, switching modalities and acquisition rates, and triggering targeted experiments. We then discuss the remaining challenges and future outlook. Overall, intelligent microscopes that can sense, analyse, and adapt promise to transform optical imaging by opening new experimental possibilities.

data-driven | reactive microscopy | image analysis | machine learning

Correspondence: *(H. S. Heil)* hsheil@igc.gulbenkian.pt, *(R. Henriques)* rjhenriques@igc.gulbenkian.pt r.henriques@ucl.ac.uk

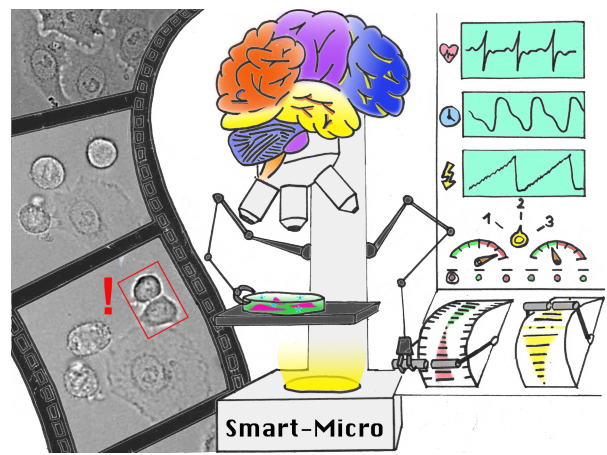

**Data-driven microscope:** The data-driven microscope integrates advanced computational techniques into its imaging capabilities. It uses machine learning algorithms and real time data analysis to automatically adjust the acquisition parameters. This way it is possible to optimise imaging conditions, enhance image quality and extract meaningful information without heavy reliance on manual intervention.

## Introduction

Optical microscopy techniques, such as brightfield, phase contrast, fluorescence, and super-resolution imaging, are widely used in life sciences to obtain valuable spatiotemporal information for studying cells and model organisms. However, these techniques have certain limitations with respect to critical parameters such as resolution, acquisition speed, signal to noise ratio, field of view, extent of multiplexing, z-depth dimensions and phototoxicity. The trade-offs between these critical imaging parameters are often represented within a "pyramid of frustration" (Fig. 1A). Although improving hardware can extend capabilities, optimal balancing depends on the imaging context. Especially, as scientific research delves into more complex questions, trying to understand the mechanisms of cell and infection biology at a molecular level in physiological context, traditional static microscopes may not be sufficient to capture relevant dynamics or rare events. Innovative efforts focus on overcoming these restrictions through integrated automation. Data-driven microscopes employ real-time data analysis to dynamically control and adapt acquisition (Fig. 1B). The core concept involves introducing automated feedback loops between image-data interpretation and microscope parameters tuning. Quantitative metrics extracted via computational analysis then dictate adaptive protocols tailored to phenomena of interest. The system reacts to predefined observational triggers by optimising imaging parameters - such as excitation, stage position, and objective lenses - to capture critical events efficiently (Fig. 1C).

Image analysis algorithms are pivotal in data-driven methodologies with customised approaches serving a large variety of situations. These approaches can use machine learning techniques to perform tasks such as classification, segmentation, tracking, and reconstruction without the need for explicit programming. By integrating machine learning, intelligent microscopes can make contextual decisions by identifying subtle features that traditional rule-based software may miss. Thus, these data-driven principles are able to increase the efficiency of image acquisition and enrich the information contents in diverse scenarios, especially in high throughput and high content imaging. It enables to capture discrete and rare events at different temporal and spatial scales and relate it to population behaviour. This information cannot be accessed with classical imaging approaches, especially because they would require extended exposure to cell damaging imaging conditions (1).

In this review, we will first introduce the concept of data-driven microscopy and the common methods used to address microscopy challenges. Then, we will explain the principles and frameworks that enable reactive machine learning-based data-driven systems. Finally, we will showcase various applications that benefit from the integration of data-driven microscopy to highlight new experimental possibilities.

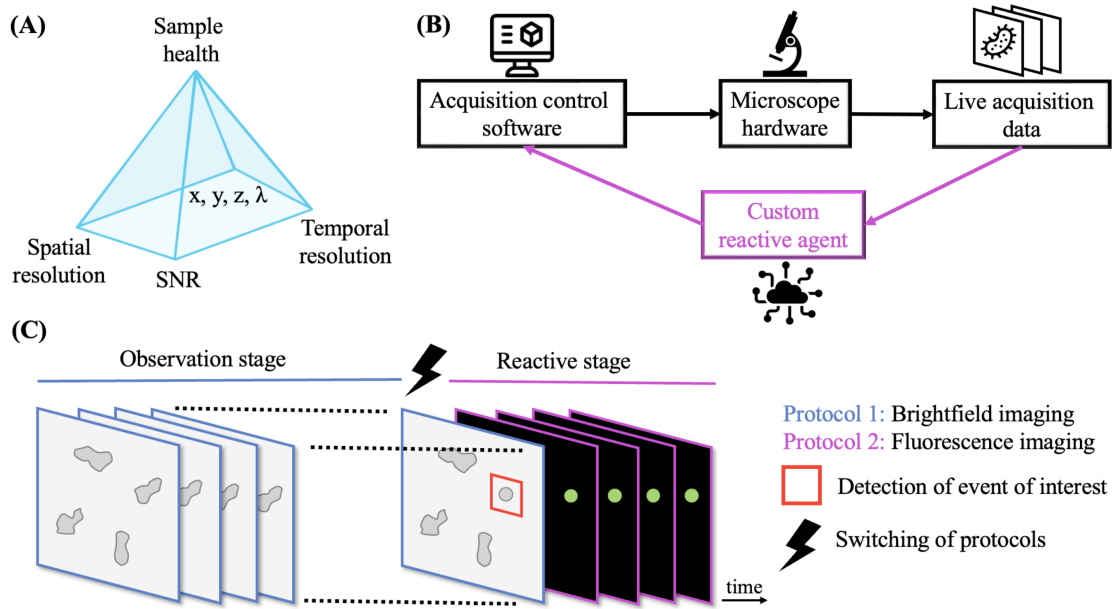

**Fig. 1. Data-driven microscopy workflow.** (A) Pyramid of frustration: Trade-offs in acquisition parameters are visualised as a pyramid, highlighting the interdependence of signal-to-noise ratio (SNR), sample health, temporal resolution, spatial resolution, and the extend to the field of view, 3D volume and multiplexing (x, y, z, λ dimensions). Enhancing one parameter typically compromises at least one other; (B) Schematic of workflow: Acquisition control software: software-driven control of microscope hardware for image capture; Microscope hardware: imaging devices with programmatic interface; Custom reactive agent: Integration of a custom reactive agent that analyses live acquisition data, providing real-time feedback to the software to adapt the acquisition parameters; (C) Acquisition stages: Observation stage (Blue): The sample is continuously monitored using a simple imaging protocol, such as brightfield. This stage is non-invasive and preserves sample health; Reactive stage (Magenta): Upon detection of a target event (e.g., a cell entering a specific cell-cycle stage), the reactive agent initiates a fluorescence imaging protocol, enabling detailed observation of the event.

## Data-Driven Microscopy

Data-driven microscopes can analyse imaging data in real-time and execute predefined actions upon specific triggers. These reactive systems feature feedback loops between quantitative image analysis and microscope control, which allows them to tailor data acquisition to objects or phenomena of interest. Specifically for event-driven approaches, this trigger can be based on detecting the occurrence of a specific event. By implementing the prediction of states of interest even smart or intelligent microscopy approaches can be realised.

A recent work showcasing event-triggered protocol switching is by Oscar André *et al.* (2). They, for example, performed dual-scale imaging of host-pathogen interactions using a co-culture model. The system first continuously scanned multiple fields of view of the sample at low magnification to monitor interactions between fluorescently labelled human cells and bacteria. An integrated algorithm analysed each frame to detect interaction events based on proximity analysis. Upon detecting a target number of cell-bacteria interactions, the system automatically switched to a higher numerical aperture objective and acquisition speed and imaged the identified interactions. This allowed to capture the cellular actin remodelling induced by the infection at high temporal-spatial resolution. This dual-scale approach balances population-level behavioural monitoring and targeted high-resolution data collection in a highly efficient and high-content manner.

In super-resolution microscopy, data-driven strategies help mitigate inherent trade-offs between resolution, speed, field of view and phototoxicity during live imaging. A system by Jonatan Alvelid *et al.* (3) combines fast widefield surveillance with precisely targeted nanoscopy imaging. For instance, cultured neurons expressing genetically encoded calcium indicators were continuously monitored with widefield imaging to detect neuronal activity. Real-time analysis of calcium dynamics allowed the detection of spike events and localisation of regions of interest. Upon spike detection, the system rapidly positioned and activated Stimulated Emission Depletion (STED) nanoscopy illumination at identified sites to visualise synaptic vesicle dynamics. By limiting high-intensity light exposure spatially and temporally only when critical events occurred, this selective super-resolution imaging approach reduced cumulative photon dose by over 75% compared to continuous STED acquisition.

Beyond adjusting microscope hardware, data-driven systems can coordinate external experimental devices by integrating microfluidics control software. An automated live-to-fixed cell imaging platform called NanoJ-Fluidics, developed by Pedro Almada *et al.* (4), performs buffer exchange directly on the microscope stage. The system uses simple epifluorescence image analysis to detect cell rounding at the onset of mitosis. Upon rounding detection, NanoJ-Fluidics triggers fixation, permeabilization, and fluorescent labelling through sequential perfusion, preparing the cells for subsequent super-resolution imaging.

These examples showcase the reliance on traditional image analysis techniques to identify events of interest, which typically involve signal colocalization, intensity, or shape thresholds. However, the integration of machine learning-based image analysis can elevate reactive data-driven microscopy to a new level by enabling the detection of subtle and complex features that would otherwise go unnoticed.

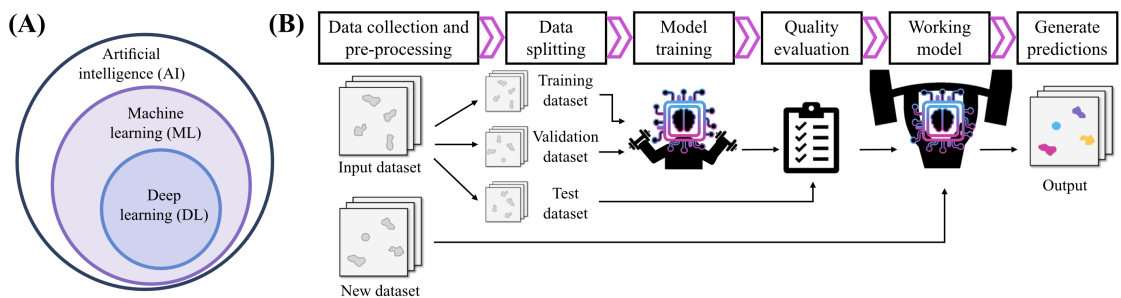

**Fig. 2. Overview of machine learning concepts for microscopy image analysis.** (A) Schematic depicting that deep learning is a subset of machine learning, which is in turn a subset of artificial intelligence. (B) Schematic of major steps on training and using a machine learning model. First, data is collected, pre-processed and split into training, validation and test datasets. Models are trained on the training dataset and training is evaluated with the validation set to prevent overfitting. Once trained, a quality control of the model is done using an independent test dataset and if positive, the model can be used to generate predictions on new unseen data.

## Machine learning for automated microscopy image analysis

Recent advances in machine learning, particularly in deep learning neural networks, have revolutionised automated image analysis for microscopy (5, 6). By training on a sufficient amount of data, machine learning models can achieve or surpass human performance in complex image processing tasks such as cell identification, structure segmentation, motion tracking, and signal or resolution enhancement. Different models excel in various aspects crucial for enhancing microscopy imaging experiments. In this section, we will introduce fundamental machine learning concepts and highlight learning strategies well-suited for microscopy imaging tasks.

Machine learning involves algorithms that learn patterns from data to make predictions without explicit programming. It falls under the umbrella of artificial intelligence, aiming to imitate intelligent behaviour (Fig. 2A). Through a training process, the algorithms tune the parameters of a specific image processing model to perform one particular task. Thus, machine learning practice requires training data, validation data and test data. The latter two dataset are used to evaluate the performance of the model during and after the training, respectively. Upon a positive quality evaluation result, the model can be used in new unseen data to make the predictions (Fig. 2B). In supervised learning, the model is trained on matched input and output examples, like images and labels, to infer general mapping functions. Unsupervised learning finds intrinsic structures within unlabelled data through techniques like clustering. As a third training category, self-supervised methods run with unlabelled data as they derive supervisory features from natural characteristics of the data itself.

A relatively simple but powerful machine learning model is the support vector machine (SVM) (7) (Figure 3). SVMs excel at classification tasks such as identifying cell types in images. SVMs plot each image as a point in a multidimensional feature space and tries to find the optimal dividing hyperplane between classes. New images are classified based on which side of the hyperplane their features fall on. SVMs have good generalisation ability provided that the features extracted from the classes are descriptive enough as to characterise them. In microscopy, SVMs are often used for initial proof-of-concept experiments to classify images into binary categories like mitotic or non-mitotic cells. Their simplicity makes SVMs convenient for implementing basic feedback loops, for instance, triggering high-resolution imaging when a specific cell type is detected.

Deep learning models including convolutional neural networks (CNNs) are state-of-the-art for complex image processing tasks. CNNs are made up of artificial neurons trained to recognise patterns in image data. One of the most influential CNN architectures is the U-Net (8) (Figure 3), which was first introduced in 2015. U-Nets have encoder layers that capture hierarchical contextual information and down sample the data, and decoder layers which rebuild the information back into a detailed map using information from the encoder path passed through the skip connections. Compared to SVMs, U-Nets can handle raw images by automatically extracting a rich feature representation, and therefore, it performs better on datasets with increased complexity. In microscopy, U-Nets excel at segmentation tasks like identifying and delineating different cell types, nuclei or components in the image. Their ability to recognise complex structures based on contextual understanding of images makes U-Nets well-suited for implementing data-driven microscopy feedback loops. For example, U-Nets could be used to alter illumination or magnification when specific cellular structures are detected.

An additional powerful class of machine learning approaches gaining traction in microscopy are generative adversarial networks (GANs) (9) (Figure 3). GANs contain paired generator and discriminator networks trained in an adversarial manner. The generator creates synthetic images to mimic real data, while the discriminator classifies images as real or fake. Competing drives the generator to produce increasingly re-alistic outputs. In microscopy, GANs are applied for data augmentation, image enhancement, modality translation, and simulation. For instance, GANs can create diverse training data, convert brightfield to fluorescence-like images, or simulate images under different conditions. A major benefit of GANs is that it can be unsupervised and thus no labelled or paired datasets are needed to train them. For smart microscopes, GANs could enable pre-processing loops to improve image quality before analysis and provide an estimate of variability or confidence in the generated results to prioritise tasks. They also hold promise for predicting nanoscale information to guide super-resolution imaging.

| Support Vector Machine (SVM) | U-Net | Generative Adversarial Network (GAN) |
|---|---|---|
| 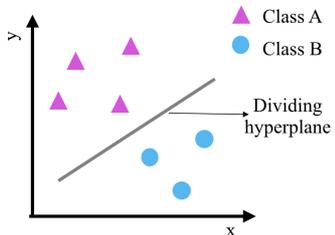 | 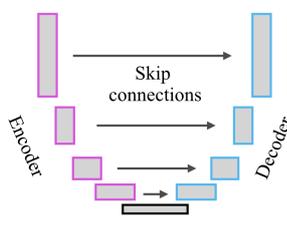 | 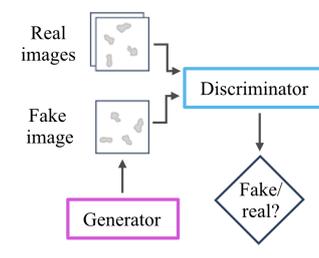 |
| **Benefits:** <br>• High generalisation ability <br>• Simple to train <br>• Easy to interpret | • Excel in contextual pattern and feature extraction without previous programming <br>• Accurate and scalable to large complex datasets <br>• Simple to implement | • Same as U-Net <br>• Generation of new high-quality synthetic images, that can be used for data augmentation <br>• No annotated data needed |
| **Limitations:** <br>• Struggle with complex datasets <br>• Poor performance on overlapping class problems | • Require large amounts of labelled training data <br>• Computationally intensive <br>• Can also become "black boxes", making it difficult to interpret their decision-making process. | • Lack of established quality assessment <br>• Prone to artificial hallucinations <br>• Computationally heavy <br>• Lack of interpretability |
| **Image analysis tasks:** Primarily used for image classification and image binary segmentation, *e.g.* distinguishing between cell types or identifying presence of specific structures. | Suitable for complex tasks such as semantic segmentation, resolution enhancement, denoising, restoration, image translation | Same as the U-Net, but without the need of labelled data. |

**Fig. 3. Comparative analysis of machine learning algorithms in data-driven microscopy.** Comparison of various machine learning (ML) algorithms employed in data-driven microscopy, delineating their respective advantages, limitations, and applications for image analysis tasks.

Machine learning provides data-driven microscopy with flexibility and empowers faster and more adaptative imaging workflows. Trained models extract relevant information from images that is then used to optimise data collection by adapting microscope parameters accordingly in real-time. This transformative potential has been demonstrated across diverse imaging modalities, as highlighted in the next section.

## Applications of machine learning powered reactive microscopy

Artificial intelligence is driving the development of intelligent microscopes that can sense, analyse, and adapt in real-time. Recent innovations have demonstrated reactive imaging systems across various modalities, ranging from widefield to super-resolution microscopy. In this section, we will discuss the key applications of machine-learning-powered reactive microscopy, highlighting the potential of these systems to revolutionise optical imaging.

MicroPilot (10), a software that provides a framework for data-driven microscopy, is one of the pioneering works in the field. The system is based on LabView and C for image analysis and can be implemented in various commercial systems. It is also compatible with Micro-Manager (11), an open-source tool widely used to control microscopes. The MicroPilot study provides different examples of how cells can be monitored in low-resolution mode, and images can be analysed using a SVM trained to classify different mitotic stages. A complex experiment is triggered once a cell in a desired stage is detected. After completion, the imaging returns to scanning mode until the next detection. To demonstrate its capacity, an experiment was conducted to study the potential role of a specific protein in the condensation of mitotic chromosomes. The experiment monitored 3T3 cells and it triggered Fluorescence Recovery After Photobleaching (FRAP) acquisitions upon identification of each prophase cell. Half of the nucleus was selectively photobleached, and the signal recovery was monitored. It is worth noting that this set of experiments was completed in just four unattended nights, generating results equivalent to what would have taken a full month for an expert user.

Building on this concept, MicroMator (12) was developed to offer an open-source toolkit for reactive microscopy using Python and Micro-Manager. It includes pre-trained U-Net models to segment yeast and bacteria cells. Researchers applied MicroMator to selectively manipulate targeted cells during live imaging. One noteworthy example involves an optogenetically-driven recombination experiment in yeast. In this experiment, genetically modified yeast cells are selected for exposure to light, triggering recombination and the subsequent expression of a protein that arrests growth together with a fluorescent protein for monitoring purposes. To generate islets of recombined yeast, MicroMator's algorithm individually selects yeast cells at a minimum distance apart, tracks them and repeatedly triggers light exposure on them, increasing the chances of recombination and, thus, the amount of relevant information in the acquired data.

In addition to enhancing data information density, the image quality can be dynamically optimised based on the sample properties. One example of this is the learned adaptive multiphoton illumination (LAMI) (13) approach, which uses a physics-informed machine learning method to estimate the optimal excitation power to maintain a good SNR across depth in multiphoton microscopy. This becomes particularly relevant for non-flat samples with varying scattering regions. Given the surface characteristics of the sample, LAMI selectively adjusts the excitation power where needed, effectively expanding the imaging volume by at least an order of magnitude, while minimising the potential for photodamage effects. The effectiveness is also demonstrated by observing immune cell responses to vaccination in a mouse lymph node with live intravital multiphoton imaging. Furthermore, LAMI significantly reduces computation time by incorporating a machine learning-based method instead of a purely physics-based approach. The computation time is reduced from approximately one second per focal time point to less than one millisecond.

In a study conducted by Suliana Manley and her team, they aimed to image mitochondria division using Structured Illumination Microscopy while minimising photodamage effects (14). To achieve this, they trained a U-Net to detect spontaneous mitochondria divisions in dual-colour images labelling mitochondria and the mitochondria-shaping dynamin-related protein 1. The model was integrated into the imaging workflow to trigger interchangeably the acquisition mode from a slow imaging rate, suitable for live-cell observation, to a faster imaging rate, enabling the collection of higher time-resolved data of mitochondrial fission. Interestingly, the similarity in the morphological characteristics and protein accumulation at fission sites allowed the network to be repurposed for detecting fission events in the bacteria *C. crescentus* without additional training. The research team quantitatively assessed and compared photobleaching decay among the slow, fast, and event-driven acquisitions. When compared to the fast mode, they observed a significant reduction in photobleaching using the event driven acquisition mode, thereby extending the duration of imaging experiments. As expected, this reduction is not as big as with the slow acquisition mode, but it comes with the benefit of capturing the event with higher temporal resolution and thus the measurement of an average smaller constriction diameters that would have been otherwise missed.

Lastly, the work of Flavie Lavoie-Cardinal's research group focuses on capturing the remodelling process of dendritic F-actin, transitioning from periodic rings to fibres, within living neurons with STED imaging (15). They monitor cells with confocal microscopy based on which synthetic STED images are generated, considerably reducing photodamage effects. For this, they employ a task-assisted generative adversarial network (TA-GAN). TA-GAN's training is strengthen by also considering the error of actin ring and fibres segmentation in the synthetic images. During acquisition, the system estimates the uncertainty of the model predicting synthetic images to decide whether to initiate a real STED image acquisition and fine-tune the generative model if needed. This allows to track actin remodelling in stimulated neurons at high accuracy and resolution. Their results illustrate that this strategy can potentially reduce the overall light exposure by a significant margin, up to 70% and importantly, they manage to acquire biologically relevant live super-resolution time lapse images for 15 minutes.

By integrating machine learning into the microscopy workflow, researchers have showcased techniques to enhance data quality and quantity while minimising phototoxicity. The applications highlighted in this section demonstrate the transformative potential of machine learning-powered microscopes across diverse imaging modalities and biological questions. As machine learning methods and computational power continue advancing, we can expect even more breakthroughs in intelligent microscopy, bringing us closer to the goal of fully automated, optimised imaging platforms that accelerate biological discovery.

## Conclusions and outlook

Data-driven microscopy has demonstrated impressive capabilities in optimising illumination, modality switching, acquisition rates, and event-triggered imaging. These approaches improve image acquisition's efficiency and information content, enabling studying dynamic biological processes across different scales. Intelligent microscopes offer new experimental possibilities, from observing rare neuronal activity at the nanoscale resolution to studying immune cell dynamics in tissues.

However, realising the full potential of data-driven microscopy requires addressing technical and practical challenges. One major limitation is the need for robust and accurate machine learning models, especially when dealing with small microscopy datasets. Expanding open-source repositories of annotated images and simulations can facilitate the development and validation of new algorithms. Additionally, incorporating unsupervised and self-supervised techniques shows promise in overcoming the scarcity of labelled data.

Another critical aspect is the design of microscope hardware optimised for data-driven imaging. Retrofitting analysis and control modules into traditional systems is common, but purpose-built instrumentation that integrates software, optics, detectors, and automation is essential. For example, spatial light modulators can enable rapid adaptable illumination for optimal signal-to-noise ratio across different samples. On the detection side, high-speed, low-noise cameras or point-scanning systems tailored for live imaging can enhance acquisition speeds.

In order to increase the use of data-driven microscopy software, it needs to be made more user-friendly and accessible. This can be achieved by creating simplified interfaces for designing and executing reactive imaging experiments, allowing non-experts to take advantage of these advanced methods. Expanding open-source platforms like Micro-Manager will encourage community contributions and drive innovation. Additionally, package managers, such as BioImage Model Zoo (16), ZeroCostDL4Mic (17), and DL4MicEverywhere (18), that facilitate the sharing and installation of pre-trained models can help overcome barriers in deploying machine learning solutions.

As data-driven microscopy moves beyond proof-of-concept studies, ensuring the robustness and reproducibility of autonomous microscopes becomes crucial. Maintaining image quality control and detecting failures during unsupervised operation for extended duration is challenging. Detailed performance benchmarking across laboratories using standardised samples can help identify best practices. While this approach can be a great asset in minimising user bias, a selection bias in decision making can still arise. Here, extensive validation of machine learning predictions and adaptive decisions is required to build trust in intelligent systems.

Data-driven microscopy represents a new era for optical imaging, overcoming inherent limitations through real-time feedback and automation. Intelligent microscopes have the potential to transform bioimaging by opening up new experimental possibilities. Pioneering applications demonstrate the ability to capture dynamics, rare events, and nanoscale architecture by optimising acquisition on-the-fly. While challenges in robustness, accessibility, and validation remain, the future looks promising for microscopes that can sense, analyse, and adapt autonomously. We envision data-driven platforms becoming ubiquitous tools that empower researchers to image smarter, not just faster. The next generation of automated intelligent microscopes will provide unprecedented spatiotemporal views into biological processes across scales, fuelling fundamental discoveries.

**ACKNOWLEDGEMENTS**
This work was supported by the Gulbenkian Foundation (LM, EGM, HSH, RH), received funding from the European Union through the Horizon Europe program (AI4LIFE project with grant agreement 101057970-AI4LIFE, and RT-SuperES project with grant agreement 101099654-RT-SuperES to R.H.) and the European Research Council (ERC) under the European Union's Horizon 2020 research and innovation programme (grant agreement No. 101001332 to R.H.). Views and opinions expressed are however those of the authors only and do not necessarily reflect those of the European Union. Neither the European Union nor the granting authority can be held responsible for them. This work was also supported by the European Molecular Biology Organization (EMBO) (Installation Grant EMBO-2020-IG-4734 to RH and the postdoctoral fellowships ALTF 499-2021 to HSH and ALTF 174-2022 to EGM), the Fundação para a Ciência e Tecnologia, Portugal (FCT fellowship CEECIND/01480/2021 to HSH), the Chan Zuckerberg Initiative Visual Proteomics Grant (vpi-0000000044 with DOI:10.37921/743590vtudfp to R.H.) and the Chan Zuckerberg Initiative DAF, an advised fund of Silicon Valley Community Foundations (Chan Zuckerberg Initiative Napari Plugin Foundations Grant Cycle 2, NP2-0000000085 granted to R.H.). R.H. also acknowledges the support of LS4FUTURE Associated Laboratory (LA/P/0087/2020). LM is kindly supported by a collaboration between Abbelight and the Integrative Biology and Biomedicine (IBB) PhD programme from Instituto Gulbenkian de Ciência.

**EXTENDED AUTHOR INFORMATION**
- Leonor Morgado: 0000-0003-4510-8456; ALeonorMorgado
- Estibaliz Gómez-de-Mariscal 0000-0003-2082-3277; gomez_mariscal
- Hannah S. Heil: 0000-0003-4279-7022; Hannah_SuperRes
- Ricardo Henriques: 0000-0002-2043-5234; HenriquesLab

**AUTHOR CONTRIBUTIONS**
L.M., H.S.H, and R.H. conceptualised the majority of the manuscript. L.M. wrote the manuscript with input from H.S.H, and R.H.. E.G-M. contributed critical comments, and conceptual suggestions to improve the manuscript. All authors reviewed and edited the manuscript.

**COMPETING FINANCIAL INTERESTS**
The authors declare no competing financial interests.